\documentclass{article}
\usepackage{spconf,amsmath,graphicx}
\usepackage{amsmath,amssymb}
\usepackage{color,amsfonts,theorem}
\usepackage{graphicx}
\usepackage[english]{babel}
\usepackage{algorithm}
\usepackage{algorithmic}
\usepackage{epstopdf}
\usepackage{appendix}
\def \T {^\top}
\def \R {\mathbb{R}}

\def \tX {\underline{ \mathbf{X}}}
\def \tY {\underline{ \mathbf{Y}}}

\def \one {\mathbf{1}}
\def \zero {\mathbf{0}}

\def \vc {\mathbf{c}}

\def \tvec {\text{vec}}

\def \MA {\mathbf{A}}
\def \MB {\mathbf{B}}
\def \MC {\mathbf{C}}

\def \MH {\mathbf{H}}
\def \MI {\mathbf{I}}

\def \MP {\mathbf{P}}

\def \MS {\mathbf{S}}

\def \MU {\mathbf{U}}

\def \MX {\mathbf{X}}
\def \MY {\mathbf{Y}}
\def \MZ {\mathbf{Z}}

\def \v0 {\mathbf{0}}
\def \bPi {\boldsymbol{\Pi}}
\def \bLambda {\boldsymbol{\Lambda}}

\def \be {\begin{equation}}
\def \ee {\end{equation}}

\def \lcp {[\kern-0.15em[}
\def \rcp {]\kern-0.15em]}
\def \R {\mathbb{R}}

\newtheorem{Theorem}{Theorem}

\theoremstyle{definition}

\ninept

\title{Hyperspectral Super-Resolution: A Coupled Nonnegative Block-Term Tensor Decomposition Approach}
\name{Guoyong Zhang$^{*\dag}$, Xiao Fu$^{\dag}$, Kejun Huang$^\ddag$, and Jun Wang$^*$ }
\address{
$^{\dag}$Oregon State University, Corvallis, OR 97331 USA\\
$^\ddag$University of Florida, Gainesville, FL 32611 USA\\
$^*$University of Electronic Science and Technology of China, Chengdu, China}

\begin{document}
%
\maketitle
\begin{abstract}
Hyperspectral super-resolution (HSR) aims at fusing a hyperspectral image (HSI) and a multispectral image (MSI) to produce a super-resolution image (SRI). Recently, a coupled tensor factorization approach was proposed  to handle this challenging problem, which admits a series of advantages over the classic matrix factorization-based methods. In particular, modeling the HSI and MSI as low-rank tensors following the {\it canonical polyadic decomposition} (CPD) model, the approach is able to provably identify the SRI, under some mild conditions. However, the latent factors in the CPD model have no physical meaning, which makes utilizing prior information of spectral images as constraints or regularizations difficult---but using such information is often important in practice, especially when the data is noisy. In this work, we propose an alternative coupled tensor decomposition approach, where the HSI and MSI are assumed to follow the {\it block-term decomposition (BTD)} model. Notably, the new method also entails identifiability of the SRI under realistic conditions. More importantly, when modeling a spectral image as a BTD tensor, the latent factors have clear physical meaning, and thus prior knowledge about spectral images can be naturally incorporated. Simulations using real hyperspectral images are employed to showcase the effectiveness of the proposed approach with nonnegativity constraints.
\end{abstract}
\section{Introduction}
In recent years, {\it hypespectral super-resolution} (HSR) has attracted a lot of attention in the remote sensing community \cite{simoes2014convex, yokoya2011coupled,wei2015hyperspectral,veganzones2015hyperspectral,kanatsoulis2018hyperspectral,wu2018hi,wu2019hyperspectral}. The problem arises because of the hardware limitations of remotely deployed sensors.
To be specific, hyperspectral sensors usually come with high spectral resolution but low spatial resolution. The nultispectral sensors have the opposite. Nevertheless, both the spatial and spectral information is critical for many downstream tasks such as material detection, classification, and landscape change tracking. The HSR task aims at fusing a pair of spatially co-registered hyperspectral image (HSI) and multispectral image (MSI) to produce a super-resolution image (SRI) that exhibits high resolution in both space and frequency.

Classic methods mostly tackle this problem from a coupled matrix factorization viewpoint  \cite{simoes2014convex, yokoya2011coupled,wei2015hyperspectral,veganzones2015hyperspectral}. To be specific, following the classic linear mixture model (LMM) of the spectral pixels, a spectral image (i.e., MSI or HSI) is modeled as a product of an endmember matrix and an abundance matrix, which capture the spectral and spatial information in the described area, respectively. Simply speaking, the coupled matrix factorization approach `extracts' the endmember matrix (or its rang space) from the HSI and the abundance matrix (or its row subspace) from the MSI, and then `assemble' the SRI using these extracted two matrices.
There also exists algorithms that do not consider explicit factorization forms, but utilize the low-rank matrix structure of the spectral images, e.g., \cite{wu2019hyperspectral}.

The matrix-based approaches are well-motivated and perform reasonably well. Since the latent factors of the matrix model have strong physical interpretation (i.e., endmembers and abundances), many structural constraints and regularizations are used to enhance performance. For example, the endmembers and abundances are all nonnegative \cite{wu2018hi,wu2019hyperspectral,wei2015fast,wei2015hyperspectral}. In addition, the abundance matrix has smooth (or, small total-variation) rows since the distribution of the materials in space is continuous \cite{simoes2014convex}. Utilizing such information has proven effective, especially when the data is noisy. The challenge, however, lies in theoretical understanding. Most of the coupled matrix factorization based approaches do not have theoretical guarantees for the identification of the SRI under their respective models. Very recently, a work \cite{li2018hyperspectral} showed that identifying the SRI under the coupled matrix factorization model is possible---under a set of somewhat restrictive conditions, e.g., that the `pure pixels' of each material exists in the HSI and that the abundance matrix of the MSI is sparse enough.

Unlike the matrix based methods, the work in \cite{kanatsoulis2018hyperspectral} took a different route. There, a coupled tensor factorization framework is proposed to handle the HSR problem. The spectral images are modeled as third-order tensors whose \emph{canonical polyadic decomposition} (CPD) is essentially unique. The recovery of the SRI can be guaranteed by jointly estimating the CPD latent factors of the HSI and MSI. Latter on, a coupled tucker decomposition approach was also proposed, which is in the same vain \cite{prevost2019coupled}. Although the works in \cite{kanatsoulis2018hyperspectral,prevost2019coupled} offer elegant identifiability guarantees  for  recovering the SRI, the price to pay is also obvious: under both the CPD and Tucker models, the latent factors of the image tensors have no physical meaning---this means that it may be hard to effectively utilize known properties of the endmembers and the abundances to enhance performance.

To circumvent this issue, in this work, we propose to employ the {\it block term decomposition} (BTD) \cite{de2008decompositions1,de2008decompositions2,de2008decompositions3} model for the spectral images.
Under BTD with rank-$(L_r,L_r,1)$ terms, the latent factors have explicit physical explanations under the widely adapted LMM---i.e., endmembers and abundances maps---if the abundance maps are approximately low-rank matrices \cite{qian2016matrix}.
We show that, under reasonable conditions, the new model also guarantees the recovery of the SRI.
More importantly, constraints and regularizations that reflect the physical properties of the endmembers and abundance maps can be naturally incorporated.
In practice, spectral signatures and abundances maps are nonnegative. Hence, in this work, we propose a coupled nonnegative BTD (CNN-BTD) model to estimate the SRI, as the first step towards a fully fledged constrained coupled tensor decomposition framework. Furthermore, we introduce a  block coordinate descent (BCD) algorithm interleaved with alternating direction method of multipliers (ADMM) to solve the coupled CNN-BTD formulation. Simulation results using real hyperspectral data show that the proposed algorithm outperform the recent work in \cite{kanatsoulis2018hyperspectral} that is based on CPD.

\section{Preliminaries and Prior Work}
\subsection{Tensor Algebra}
A third-order tensor that follows the CPD model can be expressed as follows \cite{sidiropoulos2017tensor} $$\tX(i,j,k)=\sum_{f=1}^F\MA(i,f)\MB(j,f)\MC(k,f),$$
where $F$ is the minimum number such that the above holds. In many cases, we also briefly denote it as $\tX=\lcp\MA,\MB,\MC\rcp$, where $\MA\in\R^{I\times F},\MB\in\R^{J\times F},\MC\in\R^{K\times F}$ are the three latent factors of the tensor under the CPD model. The integer $F$ is referred to as the CP rank.
The CPD model is arguably the most popular tensor model in the signal processing literature, perhaps because of its powerful identifiability properties---under very mild conditions the latent factors $\MA,\MB,\MC$ are uniquely identifiable up to trivial solutions, e.g., column scaling and permutation---see a comprehensive overview in \cite{sidiropoulos2017tensor}.

%

\subsection{Block-Term Decomposition (BTD)}

In addition to the CPD model, another very useful tensor factorization model is the \emph{block-term decomposition} (BTD) model \cite{de2008decompositions1,de2008decompositions2,de2008decompositions3}. The BTD model with rank-$(L_r,L_r,1)$ terms (or $(L_r,L_r,1)$-BTD for conciseness) of a third-order tensor can be written as
\begin{equation}
\label{BTDmodel}
\tX = \sum_{r=1}^R (\MA_r\MB_r\T)\circ\vc_r,
\end{equation}
where $\MA_r\in\mathbb{R}^{I\times L_r}$ and $\MB_r\in\mathbb{R}^{J\times L_r}$ for $r=1,\ldots,R$, $\MC=[\vc_1,\cdots,\vc_R]\in\mathbb{R}^{K\times R}$, and `$\circ$' denotes the outer product.
We further denote $\MA=[\MA_1,\cdots,\MA_R]\in\R^{I\times \sum_{r=1}^RL_r}$ and $\MB=[\MB_1,\cdots,\MB_R]\in\R^{J\times \sum_{r=1}^RL_r}$.
One nice property of BTD is that under some mild conditions, $\MA_r\MB_r\T$ and  $\vc_r$ for $r=1,\ldots,R$ are identifiable up to permutation and scaling ambiguities. 
A representative result is given as follows:
\begin{Theorem} \cite{de2008decompositions2} \label{BTD_theorem}
Let $(\MA,\MB,\MC)$ represent a BTD of $\tX$ in rank-$(L,L,1)$ terms.
Assume $(\MA,\MB,\MC)$ are drawn from certain joint absolutely continuous distributions. If $IJ\ge L^2R$ and
\[
\min\left(\bigg\lfloor\frac{I}{L}\bigg\rfloor,R\right) + \min\left(\bigg\lfloor\frac{J}{L}\bigg\rfloor,R\right)+\min(K,R)\ge 2R+2,
\]
then, $\{\MA_r\MB_r\T,\vc_r\}_{r=1}^R$ are essentially unique almost surely.
\end{Theorem}

The matricization of a BTD tensor will be used in the next section. Here we introduce some necessary definitions. The classic Khatri-Rao product (column-wise Kronecker) is defined as
\begin{align*}
&\MB_r\odot\MA_r = \\
&[\MB_r(:,1)\otimes\MA_r(:,1),\cdots,\MB_r(:,L)\otimes\MA_r(:,L)]\in\R^{IJ\times L_r},
\end{align*}
where $\otimes$ denotes the Kronecker product.
The partionwise Khatri-Rao product  between two partition matrices  is defined as \cite{de2008decompositions2}
\[
\MC\odot_p\MA = [\vc_1\otimes\MA_1,\cdots,\vc_R\otimes\MA_R]\in\R^{IK\times \sum_{r=1}^RL_r}.
\]
Using the above notations, the unfoldings of $\tX = \sum_{r=1}^R (\MA_r\MB_r^\top)\circ \vc_r$
can be expressed as
\begin{subequations}\label{Xunfold}
\begin{align}
\MX_1&=(\MC\odot_p\MB)\MA^\top,\\
\MX_2&=(\MC\odot_p\MA)\MB^\top, \\
\MX_3&=[(\MB_1\odot\MA_1)\one_{L_1},\cdots,(\MB_R\odot\MA_R)\one_{L_R}]\MC^\top, \nonumber\\
     & = [\tvec(\MA_1\MB_1^\top),\cdots,\tvec(\MA_R\MB_R^\top)]\MC^\top.
\end{align}
\end{subequations}

\section{Signal Model and  Background}
 The HSI can be represented as a third-order tensor $\tY_H\in \R^{I_H\times J_H\times K_H}$, where $I_H$ and $K_H$ denote the spatial dimensions and $K_H$ denotes the number of spectral bands.
 Similarly, the MSI can be represented by $\tY_M\in\R^{I_M\times J_M\times K_M}$. Because of the resolutions of the sensors, we have $$K_M\ll K_H$$ in general. On the other hand, an MSI has a much finer resolution in spatial domain than HSI---$I_HJ_H\ll I_MJ_M$.

The goal of the HSR task is to recover an SRI $\tY_S\in \R^{I_M\times J_M\times K_H}$ from a pair of co-registerd HSI and MSI.
Following the work in \cite{kanatsoulis2018hyperspectral}, the MSI can be modeled as the mode-3 multiplication between the SRI and  a slab selection and averaging matrix $\MP_3\in\R^{K_M\times K_H}$:
 \[\tY_M=\tY_S\times_3\MP_3.\]
To obtain the HSI from the SRI,
the spatial degradation of each SRI frontal slab $\tY_S(:,:,k)$  is modeled as a combination of circularly symmetric Gaussian blurring and downsampling. This can be expressed as the following:
\[\tY_H=\tY_S\times_1\MP_1\times_2\MP_2\]
where $\MP_1\in\R^{I_H\times I_M}$, $\MP_2\in\R^{J_H\times J_M}$ are the blurring and downsampling matrices from the $x$-axis and $y$-axis in space, respectively.

If we model ${\bf  Y}_S = \lcp\MA,\MB,\MC\rcp$, a key observation in \cite{kanatsoulis2018hyperspectral} is that
\[     \tY_M =   \lcp\MA,\MB,{\bf P}_3\MC\rcp, ~   \tY_H=   \lcp {\bf P}_1 \MA, {\bf P}_2\MB,\MC\rcp .\]
A novel coupled CPD model was proposed to handle the HSR task in \cite{kanatsoulis2018hyperspectral}. The main idea is to make factors $\MA,\MB,\MC$ to fit both $\tY_M=\tY_S\times_3\MP_3$ and  $\tY_H=\tY_S\times_1\MP_1\times_2\MP_2$ in the least squares sense as follows:
\begin{small}
\[
\text{(C1)}\;\;\min_{\MA,\MB,\MC} \Vert\tY_H-\lcp\MP_1\MA,\MP_2\MB,\MC\rcp\Vert_F^2
+\Vert\tY_M-\lcp\MA,\MB,\MP_3\MC\rcp\Vert_F^2
\]
\end{small}
After estimating $\MA$, $\MB$, and $\MC$ via tackling the above, the SRI can be recovered as $\widehat{\tY}_S=\lcp \widehat{\MA},\widehat{\MB},\widehat{\MC}\rcp$. This method is effective, and comes with identifiability guarantees of the SRI. Nevertheless, since the ${\bf A}, {\bf B}$ and ${\bf C}$ do not admit any physical meaning under the CPD model of HSI/MSI, it is not easy to incorporate constraints and regularizations that normally would enhance performance, as seen in the coupled matrix decomposition cases.
The work in \cite{prevost2019coupled} that employs a coupled Tucker decomposition for HSR has the same issue.


\section{Proposed Approach}
In this work, we propose a coupled BTD based HSR fromulation.
Let $\tY_S\in \R^{I_M\times J_M\times K_H}$ be the target SRI we aim to estimate. Under the classic LMM, the mode-3 matricization $\MY_{S}^{(3)}$ can be represented as \cite{ma2013signal}:
\begin{equation}\label{eq:LMM}
\MY_{S}^{(3)}=\MS\MC\T\in\R^{I_MJ_M\times K_H},
\end{equation}
where $\MC\in\R^{K_H\times R}$ is the matrix containing the spectral signatures of $R\ll \min\{I_MJ_M,K_H\}$ endmembers and $\MS\in\R^{I_MJ_M\times R}$ is the abundance matrix. The abundance map of endmember $r$ is $\MS_r=\text{mat}(\MS(:,r))$, where the mat$(\cdot)$ operation refers to reshape a vector with size $I_MJ_M\times 1$ to an $I_M\times J_M$ matrix.

In this paper, we assume that each abundance map $\MS_r\in\R^{I_M\times J_M}$ can be approximated by a low-rank with rank-$L_r$, i.e., $$\MS_r\approx\MA_r\MB_r\T,$$ where $\MA_r\in\R^{I\times L_r},\MB_r\in\R^{J\times L_r}$.
Low-rank approximation of $\MS_r$ is very reasonable, since the spatial distribution of a material is not random.
Due to the correlation along the row and column dimensions, $\MS_r$ is likely to be low-rank.
For simplicity, let $L_r=L$ for $r=1,\cdots,R$. Reshaping $\MY_{S}^{(3)}=\MS\MC\T$ to tensor representation, the tensor $\tY_S$ admits a BTD model
\[
\tY_S=\sum_{r=1}^R (\MA_r\MB_r\T)\circ \vc_r.
\]
We aim to estimate the latent factors $\MA,\MB$ and $\MC$ from HSI $\tY_H$ and MSI $\tY_M$.
The main idea is to make factors $\MA,\MB,\MC$ to fit both $\tY_M=\tY_S\times_3\MP_3$ and  $\tY_H=\tY_S\times_1\MP_1\times_2\MP_2$.
All $\MP_1,\MP_2,\MP_3$ are assumed known. This is possible since we have the BTD representations of the model product as follows:
\begin{subequations}\label{eq:BTDmodel}
	\begin{align}
	\tY_H&=\sum_{r=1}^R (\MP_1\MA_r(\MP_2\MB_r)\T)\circ \vc_r\, \label{eq:BTDmodelA}\\
	\tY_M&=\sum_{r=1}^R (\MA_r\MB_r\T)\circ \MP_3\vc_r.
	\end{align}
\end{subequations}
Ideally, one can identify $\MC$ from $\tY_H$ and $\MA_r\MB_r\T$ from $\tY_M$, fix the ambiguities, and assemble the SRI. However, this would require that both $\tY_H$ and $\tY_M$ are identifiable BTD models. In what follows, we will show that the SRI is identifiable under much  milder conditions.

The remark here is that  under the BTD model, $\MA_r\MB_1\T$ and ${\bf c}_r$ both have nice physical interpretations. The former represents the abundance map of material $r$, while the latter the spectral signature of material $r$ (i.e., the $r$th endmember). This is very different from the models under CPD and Tucker in \cite{kanatsoulis2018hyperspectral,prevost2019coupled}.

Before we go to the next subsection, we should mention that using the $(L_r,L_r,1)$-BTD to model a spectral image was first considered in \cite{qian2016matrix} for hyperspectral  unmixing. But utilizing this model for HSR was never considered, to the best knowledge of the authors. Nevertheless, the interesting results obtained in \cite{qian2016matrix} offers numerical evidence for that real spectral images can be well approximated by the $(L_r,L_r,1)$ BTD model.

\subsection{Identifiability}
In practice, MSI is likely to satisfy  the condition of Theorem~\ref{BTD_theorem} since it admits large $I_M$ and $J_M$. If the MSI is identifiable, then the $\MA_r\MB_r^{\T}$'s can be identified via applying BTD to MSI. Subsequently, $\MC$ can be estimated from mode-3 matricization of HSI, namely $\MY_{H}^{(3)}$, if $I_HJ_H\ge R$, which can be easily satisfied since in practical $R$ is very small  (i.e., $R$ is the number of materials contained in the area captured by the HSI and MSI). Based on the above intuition, we have the following identifiability theorem for coupled BTD model:

\begin{Theorem} \label{Cp_BTD_theorem}
Let $(\MA,\MB,\MC)$ represent a BTD of $\tY_S$ in rank-$(L,L,1)$ terms. Let  $\tY_M=\tY_S\times_3\MP_3$ and $\tY_H=\tY_S\times_1\MP_1\times_2\MP_2$.
Assume $(\MA,\MB,\MC)$ are drawn from certain joint absolutely continuous distributions. If $I_MJ_M\ge L^2R,I_HJ_H\ge R$, and
\[
\min\left(\bigg\lfloor\frac{I_M}{L}\bigg\rfloor,R\right) + \min\left(\bigg\lfloor\frac{J_M}{L}\bigg\rfloor,R\right)+\min(K_M,R)\ge 2R+2,
\]
then, $\{\MA_r\MB_r\T,\vc_r\}_{r=1}^R$ are essentially unique almost surely.
\end{Theorem}

The proof is relegated to Appendix A. Theorem~\ref{Cp_BTD_theorem} indicates that if the number of materials contained in the spectral images is not very large, and if tthe spatial distribution of the materials is smooth (so that the rank of the abundance map is not large), then the SRI is identifiable under the model in \eqref{eq:BTDmodel}. More importantly, this modeling strategy entails us the convenience for utilizing domain prior information on the endmembers and the abundance maps.


\subsection{Algorithm}
To validate Theorem~\ref{Cp_BTD_theorem} and to showcase the usefulness of incorporating prior information,
we propose an algorithm for decomposing the model in \eqref{eq:BTDmodel} with nonnegativity constraints on the latent factors. Nonnegativity is natural here since both the spectral signatures and their abundances are nonegative.
To this end, we propose the following problem formulation under the aforementioned coupled nonnegative BTD (CNN-BTD) model:
\begin{align*}
\text{(B1)}\;\;\min_{\MA,\MB,\MC} \quad &\Vert\tY_H-\sum_{r=1}^R (\MP_1\MA_r(\MP_2\MB_r)\T)\circ \vc_r\Vert_F^2 \\
&\qquad+ \Vert\tY_M-\sum_{r=1}^R (\MA_r\MB_r\T)\circ \MP_3\vc_r\Vert_F^2\\
\text{subject to}\;\;&\MA\ge\zero,\MB\ge\zero,\MC\ge\zero
\end{align*}
We denote the objective function of (B1) as $J(\MA,\MB,\MC)$.

We propose to employ the \textit{block coordinate descent} (BCD) scheme for handling the above problem---i.e., we update $\MA,\MB,\MC$ via solving subproblems w.r.t. each one of them while fixing the others, in a cyclical fashion. To be more specific, we update $\MA,\MB,\MC$ as below:
\begin{subequations}\label{eq:BCD}
	\begin{align}
	 \MA^{t+1} &\leftarrow \arg\min_{\MA\ge\zero} J(\MA,\MB^t,\MC^t)\\
	 \MB^{t+1} &\leftarrow \arg\min_{\MB\ge\zero} J(\MA^{t+1},\MB,\MC^t)\\
	  \MC^{t+1} &\leftarrow \arg\min_{\MC\ge\zero} J(\MA^{t+1},\MB^{t+1},\MC)
	\end{align}
\end{subequations}
where $t$ is the iteration index.

To proceed, consider the subproblem (\ref{eq:BCD}a). Using matricization, one can readily see that the subproblem is a constrained quadratic program as follows:
\begin{align}\label{Aproblem}
\MA\leftarrow\arg\min_{\MA\ge\zero}&\Vert\MY_{H}^{(1)}-(\MC\odot_p\MP_2\MB)(\MP_1\MA)\T\Vert_F^2 \nonumber\\
&+ \Vert\MY_{M}^{(1)}-(\MP_3\MC\odot_p\MB)\MA\T\Vert_F^2
\end{align}
where $\MY_{H}^{(1)}$ and $\MY_{M}^{(1)}$ denote the mode-$1$ matricization of $\tY_H$ and $\tY_M$, respectively.

We apply the alternating direction method of multipliers (ADMM) \cite{boyd2011distributed,huang2016flexible} to solve (\ref{Aproblem}). The procedure is shown in Algorithm 1. The remaining subproblem (\ref{eq:BCD}b) and (\ref{eq:BCD}c) can be handled in the same way as (\ref{eq:BCD}a).

\begin{algorithm}[htb]
\caption{Solve (\ref{Aproblem}) using ADMM}
\begin{algorithmic}
\STATE \textbf{Input}: $\MB,\MC,\MP_1,\MP_2,\rho,\MY_H^{(1)},\MY_M^{(1)}$,\texttt{Inner\_Iter}
\STATE \textbf{Initialization:} initialize $\MZ$ and $\MU$
\STATE $\MH_1=\MP_1\T\MP_1$,$\MH_3=\MI_I$
\STATE $\MH_2=(\MC\odot_p\MP_2\MB)\T(\MC\odot_p\MP_2\MB)$
\STATE $\MH_4=(\MP_3\MC\odot_p\MB)\T(\MP_3\MC\odot_p\MB)+\rho\MI_R$
\STATE $\tilde{\MH}_5=\MP_1\T\MY_H^{(1)\intercal}(\MC\odot_p\MP_2\MB) + \MY_M^{(1)\intercal}(\MP_3\MC\odot_p\MB)$
\WHILE{iteration $\le$ \texttt{Inner\_Iter}}
\STATE $\MH_5=\tilde{\MH}_5 +\rho(\MZ+\MU) $
\STATE $\MA\leftarrow$ solve Sylvester equation \cite{gardiner1992solution} $\MH_1\MA\MH_2+\MH_3\MA\MH_4=\MH_5$
\STATE update $\MZ\leftarrow[\MA-\MU]_+$, $[\cdot]_+$ is zeroing out the negative value.
\STATE update $\MU\leftarrow\MU+ (\MZ-\MA)$
\ENDWHILE
\STATE \textbf{Output:}   $\MA$
\end{algorithmic}
\end{algorithm}

\section{Simulation Results}
In this section, we provide numerical results to validate the proposed algorithms.
Following the simulation method in the literature \cite{wu2019hyperspectral,wu2018hi,kanatsoulis2018hyperspectral},
we use a publicly available HSI to act as the SRI, and synthetically generate  HSI and MSI in a realistic manner, following the so-called Wald's protocol \cite{wald1997fusion}.
The degradation from SRI to HSI is modeled as follows:The SRI is blurred by a $9\times 9$ Gaussian kernel and the downsampled 1 out of every $5\times 5=25$ pixels. The spectral response $\MP_3$ follows the setting as in \cite{kanatsoulis2018hyperspectral}. Zero-mean i.i.d. Gaussian noise is added to both HSI and MSI. In this section, the SNR is set to be 30dB.

\begin{figure}
  \centering
  \includegraphics[width=.9\linewidth]{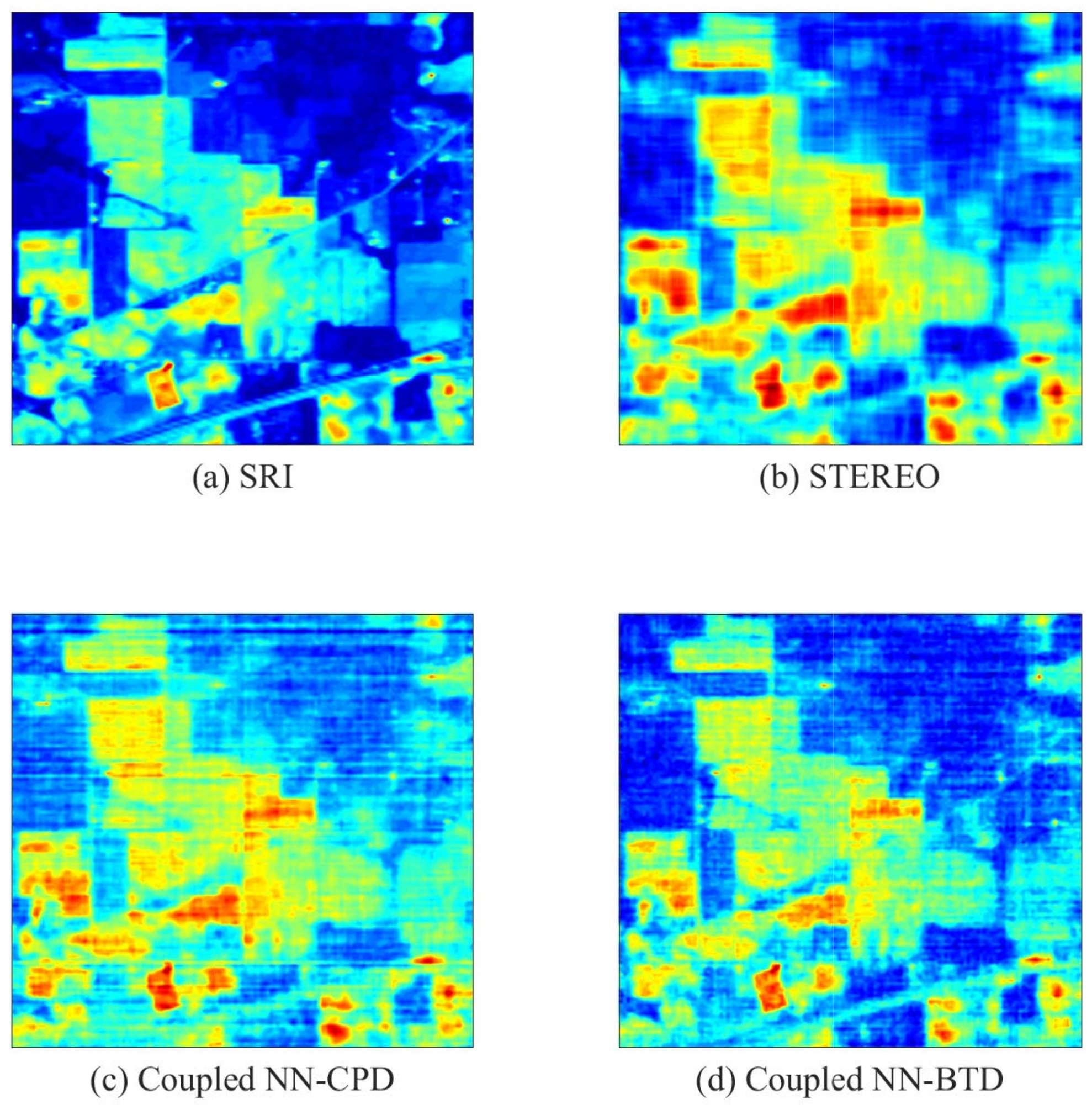}
  \vspace{-12pt}
  \caption{Indian Pines reconstruction, the 20-th band.}\label{Fig1}
\end{figure}

The baseline algorithm used for comparison is the coupled CPD approach proposed in \cite{kanatsoulis2018hyperspectral} (i.e.,\texttt{STEREO}. Besides, we also provide a nonnegativity-constrained version of \texttt{STEREO}, denoted as coupled nonnegative CPD (CNN-CPD). The CNN-CPD aim to solve Problem (C1) with additional nonnegative constraints on $\MA,\MB,\MC$.  Since the factors under the CPD model have no physical interpretation, we use this heuristic to observe if physical meaning-driven modeling and constraint-imposing are beneficial.
We also use BCD and ADMM to solve the CNN-CPD problem, whose procedure is the same as our proposed CNN-BTD. All the algorithms are handled by BCD. The number of iterations of BCD for \texttt{STEREO} are set to 100, while that of coupled NN-CPD and coupled NN-BTD are set to be 20. The number of iterations (namely \texttt{Inner\_Iter})  of each subproblem  with nonnegative constraint are set to be 5, which is to keep the same  number of solving Sylvester equation. For  CNN-CPD methods, tensor rank $F$ are set to $F=100$. For CNN-BTD, we let $R=10$ and $L=20$.

To evaluate the performance of the algorithms, We use the reconstruction SNR (\texttt{R-SNR}), which is defined as $10\log(\Vert\tY_S\Vert_F^2/\Vert\tY_S-\hat{\tY}_S\Vert_F^2)$, along with cross correlation (\texttt{CC}), spectral angle mapper (\texttt{SAM}) and relative dimensional global error (\texttt{ERGAS}) defined in \cite{kanatsoulis2018hyperspectral,loncan2015hyperspectral}. Briefly, the high \texttt{R-SNR} and \texttt{CC} values, and the low \texttt{SAM} and \texttt{ERGAS} values indicate good super-resolution performance.
\begin{small}
\begin{table}\scriptsize
\caption{Performance of the algorithms in Indian Pines data}
\begin{center}
\begin{tabular}{ |c|c|c|c|c|c| }
\hline
Algorithm&	R-SNR	&CC&	SAM	&ERGAS&runtime(sec)\\ \hline
STEREO&	25.38&	0.7841&	0.0382  &	0.9251&31.14\\ \hline
CNN-CPD&	23.2069&	0.7216&	0.0583&	1.3917&62.61\\ \hline
CNN-BTD&\textbf{27.17}&	\textbf{0.8045}&	\textbf{0.0352}&	\textbf{0.8641}&\textbf{26.64}\\  \hline
\end{tabular}
\end{center}
\end{table}
\end{small}

We use the Indian Pines HSI downloaded from the AVIRIS platform to act as the SRI. The HSI $\tY_H\in\R^{29\times 29\times 220}$ and MSI $\tY_H\in\R^{145\times 145\times 4}$ are generated from the SRI $\tY_S\in\R^{145\times 145\times 220}$ as described. Table 1 shows \texttt{R-SNR}, \texttt{CC}, \texttt{SAM}, \texttt{ERGAS} performance metrics and running times of the three algorithms averaged over 10 Monte Carlo simulations. One can see that the proposed CNN-BTD has the best super-resolution performance and the fastest speed under the tested setting, while the CNN-CPD is not as promising.

Figure. \ref{Fig1}  shows the ground-truth SRI $\tY_S(:,:,20)$ and estimated $\hat{\tY}_S(:,:,20)$ of the algorithms, as a visual reference.

\section{Conclusions}
In this work, we proposed a block term decomposition-based model for hyperspectral super-resolution. Our  motivation is to come up with an identifiability guaranteed HSR model with physical properties of the image data taken into consideration.
This way, effective regularizations and constraints can be utilized in practice, to fend against challenging scenarios, e.g., where the noise or modeling error level is high.
The proposed method can clearly serve both purposes. We also tested the proposed algorithm on a semi-real data, which shows promising results. In the future, we will consider 1) more types of regularizations and constraints that are widely adopted in hyperspectral imaging, e.g., smoothness of the endmembers and total variation of the abundance maps; and 2) testing the algorithms under many more scenarios and using a variety of different datasets.

\bibliographystyle{IEEEbib}
\bibliography{CAMSAP_Guoyong}
\clearpage
\appendices
\section{Proof of Theorem 2}
Let $\{\MA_r\MB_r\T,\vc_r\}_{r=1}^R$ (or $(\MA,\MB,\MC)$) denote the ground-truth factors and $\{\widehat{\MA}_r\widehat{\MB}_r\T,\widehat{\vc}_r\}_{r=1}^R$ (or $(\widehat{\MA},\widehat{\MB},\widehat{\MC})$) denote the estimated factors. We aim to prove that $\{\widehat{\MA}_r\widehat{\MB}_r\T,\widehat{\vc}_r\}_{r=1}^R$  is essentially the ground-truth $\{\MA_r\MB_r\T,\vc_r\}_{r=1}^R$  up to permutation and scaling ambiguities.

Denote $\MS_r=\MA_r\MB_r\T$ and $\widehat{\MS}_r=\widehat{\MA}_r\widehat{\MB}_r\T$. Let
\begin{align*}
\MS &= [\tvec(\MS_1),\cdots,\tvec(\MS_R)]\\
\widehat{\MS} &= [\tvec(\widehat{\MS}_1),\cdots,\tvec(\widehat{\MS}_R)]
\end{align*}
Note that the conditions  $I_MJ_M\ge L^2R$ and
\[
\min\left(\bigg\lfloor\frac{I_M}{L}\bigg\rfloor,R\right) + \min\left(\bigg\lfloor\frac{J_M}{L}\bigg\rfloor,R\right)+\min(K_M,R)\ge 2R+2,
\]
hold. By  Theorem 1, $\{\MA_r\MB_r\T,\vc_r\}_{r=1}^R$ can be identified from $\tY_M$ up to scaling and permutation ambiguities. Let $\{\widehat{\MA}_r,\widehat{\MB}_r\}_{r=1}^R$ denote the estimated factors by performing BTD of $\tY_M$. Therefore, $\{\widehat{\MS}_r\}_{r=1}^R$ is a scaling and permutation version of ground-truth $\{\MS_r\}_{r=1}^R$ and we have
\[
\widehat{\MS}=\MS\bPi\bLambda,
\]
where $\bPi$ is a permutation matrix and $\bLambda$ is a nonsingular diagonal matrix.

Consider the mode-3 matricization of $\tY_H$ in (\ref{eq:BTDmodelA}):
\begin{align*}
\MY_H^{(3)}&=[\cdots,\tvec(\MP_1\MA_r\MB_r\T\MP_2\T),\cdots]\MC\T\\
&=[\cdots,(\MP_2\otimes\MP_1)\tvec(\MA_r\MB_r\T),\cdots]\MC\T\\
& =(\MP_2\otimes\MP_1)\MS\MC\T\\
& = (\MP_2\otimes\MP_1)\MS\bPi\bLambda\bLambda^{-1}\bPi\T\MC\T\\
& = (\MP_2\otimes\MP_1)\widehat{\MS}(\MC\bPi\bLambda^{-1})\T.
\end{align*}
Since $(\MP_2\otimes\MP_1)\widehat{\MS}\in\R^{I_HJ_H\times R}$ and $I_HJ_H\ge R$,  $(\MP_2\otimes\MP_1)\widehat{\MS}$ is full column rank almost surely. We estimate $\widehat{\MC}$ from equation
\[
\MY_H^{(3)} =  (\MP_2\otimes\MP_1)\widehat{\MS}\widehat{\MC}\T.
\]
The solution is unique and satisfies $\widehat{\MC}=\MC\bPi\bLambda^{-1}$.

Therefore, we have proven that the estimated $(\widehat{\MC},\widehat{\MS})$ is essentially the ground-truth
$(\MC,\MS)$ up to permutation and scaling ambiguities.

\end{document}